\documentclass{article}
\usepackage{latexsym}
\usepackage{amsmath}
\usepackage{amsfonts}
\usepackage{amssymb}

\newcommand{\ms}{\medskip}

\newcommand{\noi}{\noindent}
\newcommand{\ra}{\rightarrow}
\newcommand{\bea}{\begin{eqnarray}}
\newcommand{\eea}{\end{eqnarray}}
\newcommand{\ol}{\overline}
\newcommand{\gr}{Groenewold}
\newcommand{\vh}{Van~Hove}
\newcommand{\vn}{Von~Neumann}
\newcommand{\q}{{\cal Q}}
\newcommand{\p}{C^{\infty}(M)}

\newcommand{\ci}{C^{\infty}}
\newcommand{\h}{{\mathcal H}}

\newcommand{\A}{{\mathcal A}}

\newcommand{\oo}{{\mathcal O}}
\newcommand{\s}{{\mathcal S}}

\newcommand{\fd}{finite-dimensional}
\newcommand{\id}{infinite-dimensional}

\newcommand{\pa}{Poisson algebra}
\newcommand{\pb}{Poisson bracket}
\newcommand{\la}{Lie algebra}
\newcommand{\lsa}{Lie subalgebra}
\newcommand{\sa}{subalgebra}
\newcommand{\sm}{symplectic manifold}
\newcommand{\ba}{basic algebra}
\newcommand{\fb}{{\mathfrak b}}

\newcommand{\T}{{T^*\!\,}}

\def\endproof{\hfill $\Box$}

\def\bc{{\bf C}}
\def\r{{\bf R}}

\newtheorem{thm}{Theorem}
\newtheorem{lem}{Lemma}
\newtheorem{cor}[thm]{Corollary}
\newtheorem{prop}[thm]{Proposition}

\def\sp{{\rm span}}

\begin{document}


\title{On quantizing nilpotent and solvable \ba s}

\author{{\bf Mark J. Gotay}\thanks{Supported in part by NSF grant
96-23083. E-mail: gotay@math.hawaii.edu \ \ Home Page:
www.math.hawaii.edu/\~{}gotay} \\
Department of Mathematics \\ University of Hawai`i \\ 2565 The
Mall \\ Honolulu, HI 96822  USA 
\and {\bf Janusz Grabowski}\thanks{Supported by KBN, grant No. 2 PO3A
042 10. E-mail: jagrab@mimuw.edu.pl} \\ Institute of
Mathematics \\ University of Warsaw \\ ul. Banacha 2 \\ 02-097
Warsaw, Poland}

\date{January 25, 1999 \\ (Revised March 12, 1999)}

\maketitle


\begin{abstract}
We prove an algebraic ``no-go theorem'' to the effect that a
nontrivial \pa\ cannot be realized as an associative
algebra with the commu\-ta\-tor bracket. Using this, we
show that there is an obstruction to 
quantizing the \pa\ of polynomials generated by a nilpotent
\ba\  on a \sm. Finally, we explicitly construct a polynomial
quantization of a \sm\ with a solvable \ba, thereby showing that the
obstruction in the nilpotent case does not extend to the solvable
case.
\end{abstract}
 

\begin{section}{Introduction}

We continue our study of \gr -\vh\ obstructions to
quantization. Let $M$ be a \sm,
and suppose that $\fb$ is a \fd\ ``\ba '' of observables on $M$.
Given a \lsa\ $\oo$ of the \pa\ $\p$ containing $\fb$, we are
interested in determining whether the pair $(\oo,\fb)$ can
be ``quantized.'' (See \S 2 for the precise definitions.)
Already we know that such
obstructions exist in many circumstances: In \cite{GGG} we
showed that there are no nontrivial quantizations of the pair
$(P(\fb),\fb)$ on a compact symplectic manifold, where $P(\fb)$ is the
\pa\ of polynomials on $M$ generated by $\fb$. Furthermore, in
\cite{GG2} we proved that there are no nontrivial \fd\ quantizations
of $(\oo,\fb)$ on a noncompact symplectic manifold, for any such \sa\
$\oo.$

It remains to understand the case when
$M$ is noncompact and the quantizations are \id, which is naturally
the most interesting and difficult one. Here one has little control
over either the types of \ba s that can appear (in examples they
range from nilpotent to semisimple), their representations, or the
structure of the polynomial algebras they generate \cite{go98}.

In this paper we consider the problem of quantizing $(P(\fb),\fb)$ when
the \ba\ is nilpotent. Our main result is (\S 5):
\begin{thm}
Let $\,\fb$ be a nilpotent \ba\ on a connect\-ed symplectic
manifold. Then there is no quantization of $(P(\fb),\fb)$.
\label{thm:nil}
\end{thm}

This in
turn is a consequence of an algebraic ``no-go theorem'' to the effect
that a nontrivial \pa\ cannot be realized as an associative algebra
with the commu\-ta\-tor bracket. The latter result, which is of
independent general interest, is presented in \S 3.

When $M = \r^{2n}$ and $\fb$ is the Heisenberg algebra h($2n$),
Theorem~\ref{thm:nil} provides an entirely new proof of the classical
theorem of \gr\ \cite{Gro,go-r2n}:
\begin{cor}
There is no
quantization of the pair $\big(P({\rm h}(2n)),{\rm h}(2n)\big)$.
\label{cor:gr}
\end{cor}

We remark that this version of
the no-go theorem for $\r^{2n}$ does not use the Stone-\vn\ theorem.

A natural question is whether  this obstruction to quantization when
$\fb$ is nilpotent extends to the case when $\fb$ is solvable. We
show that it does \emph{not}; in \S 6 we explicitly construct a
polynomial quantization of $\T \r_+$ with the ``affine'' \ba\
a(1).

\end{section}


\begin{section}{Background}

Let $M$ be a connected symplectic manifold. A key ingredient in the
quantization process is the choice of a  \emph{basic algebra of
observables} in the Poisson algebra $C^\infty(M)$. This is
a \lsa\ ${\fb}$ of $C^\infty(M)$ such that:
\begin{description}
\item \rule{0mm}{0mm}
\begin{enumerate}
\vspace{-4ex}
\item[(B1)] $\fb$ is finitely generated,
\vskip 6pt
\item[(B2)] the Hamiltonian vector fields
$X_b,{b\in\fb}$, are complete, 
\vskip 6pt
\item[(B3)] $\fb$ is transitive and separating, and
\vskip 6pt
\item [(B4)] $\fb$ is a minimal \la\ satisfying these requirements.
\end{enumerate} \end{description}

\noi A subset $\fb \subset C^\infty(M)$ is ``transitive'' if
$\{X_b(m)\,|\,b\in\fb\}$ spans
$T_mM$ at every point. It is ``separating'' provided its
elements globally separate points of $M$. 

Now fix a
basic algebra
${\fb}$, and let
${\cal O}$ be any Lie subalgebra of $C^\infty(M)$ containing $1$
and
${\fb}$. Then by a \emph{quantization} of the pair $({\cal O},{\fb})$
we mean a linear map
$\q$ from $\oo$ to the linear space Op($D$) of symmetric operators
which preserve a fixed dense domain $D$ in some separable Hilbert space
$\h$, such that for all $f,g \in \oo$,

\begin{description}
\item \rule{0mm}{0mm}
\begin{enumerate}
\vspace{-4.5ex}
\item[(Q1)] ${\cal Q}(\{f,g\}) =
\frac{i}{\hbar}[{\q}(f),{\q}(g)]$,
\vskip 6pt
\item[(Q2)]  ${\cal Q}(1) = I$, 
\vskip 6pt
\item[(Q3)] if the Hamiltonian vector field $X_f$ of $f$ is complete,
then $\q(f)$ is essentially self-adjoint on $D$,
\vskip 6pt
\item[(Q4)] $\q(\fb)$ is irreducible,  
\vskip 6pt
\item[(Q5)] $D$ contains a dense set of separately analytic vectors
for some set of Lie generators of $\q(\fb),$ and
\vskip 6pt
\item[(Q6)] $\q$ represents $\fb$ faithfully.
\label{def:q}
\end{enumerate}
\end{description}

\noi Here $\{\cdot,\cdot\}$ is the \pb\ and $\hbar$ is Planck's
reduced constant. 

In this paper we are interested in ``polynomial quantizations,'' i.e.
quantizations of $(P(\fb),\fb)$.

We refer the reader to \cite{go98} for an extensive discussion of
these definitions. However, we wish to elaborate on (Q4). There we
mean irreducible in the analytic sense, viz.\ the only bounded
operators which strongly commute with all $\q(b)
\in \q(\fb)$ are scalar multiples of the identity. There is another
notion of irreducibility which is useful for our purposes: We say that
$\q(\fb)$ is \emph{algebraically irreducible} provided the
only operators in Op$(D)$ which (weakly) commute with all $\q(b)
\in \q(\fb)$ are scalar multiples of the identity. It turns out that
a quantization is automatically algebraically irreducible.
\begin{prop}
Let $\q$ be a representation of a \fd\ \la\ $\,\fb$ by symmetric
operators on an invariant dense domain
$D$ in a separable Hilbert space $\h$. If
$\,\q$ satisfies {\rm (Q4)} and {\rm (Q5)}, then
$\q(\fb)$ is alge\-bra\-i\-cally irreducible.
\label{prop:alg}
\end{prop}

\noi \emph{Proof}. We need the
following two technical results, which are proven in \cite{go-r2n}.
Denote the closure of an operator $R$ by $\bar R$.  
\begin{lem} Let $R$
be an essentially self-adjoint operator and $S$ a
closable operator which have a common dense invariant domain
$D$. Suppose that $D$ consists of analytic vectors for $R$,
and that $R$ {\rm (}weakly{\rm )} commutes with $S$. Then $\exp(i
\bar R)$ {\rm (}weakly{\rm )} commutes with $\bar S$ on $D$.
\label{lem:fa1} 
\end{lem}
\begin{lem} Let $S$ be a closable operator. If a bounded operator
{\rm (}weakly{\rm )} commutes with
$\bar S$ on
$D(S)$, then they also commute on $D(\bar S)$.
\label{lem:fa2} 
\end{lem}

By virtue of (Q5) and Corollary 1 and Theorem 3 of \cite{f-s}, we may
assume that there is a dense space $D_\omega \subseteq D$ of
separately analytic vectors for some basis 
$\,{\mathcal B} = \{B_1,\ldots,B_K\}$ of $\q(\fb)$. Suppose $T \in
{\rm Op}(D)$ (weakly) commutes with every $B_k$. 
According to \cite[Prop.~1]{f-s},  $T$ leaves
$D_\omega$ invariant. Now by \cite[\S X.6, Cor.~2]{ReSi} each
$B_k\!\restriction\! D_\omega$ is essentially self-adjoint; moreover,
$T_\omega := T \!\restriction\! D_\omega$ is symmetric and hence
closable. Upon taking $R = B_k\!\restriction\! D_\omega$ and $S =
T_\omega$ in Lemma~\ref{lem:fa1}, it follows that
$\exp(i\ol{B_k\!\restriction\! D_\omega}) = \exp(i\ol{B_k})$ and
$\ol{T_\omega}$ commute on
$D_\omega$. Lemma~\ref{lem:fa2} then shows that $\exp(i\ol{B_k})$
and $\ol{T_\omega}$ commute on $D(\ol{T_\omega})$ for all $B_k \in
\mathcal B$. 

By (Q5) the representation $\q$ of $\fb$ can be integrated to a
unitary representation $\mathfrak Q$ of the corresponding
connected, simply connected group $G$ on $\mathcal H$
\cite[Cor.~1]{f-s} which, according to (Q4), is irreducible. {}From
the construction of coordinates of the second kind on $\mathfrak Q(G)$, the
map $\r^K \ra \mathfrak Q(G)$ given by
\[(t_1,\ldots,t_K) \mapsto \exp(it_1 \ol {B_1})\cdots
\exp(it_K \ol {B_K})\] 

\noi is a diffeomorphism of an open neighborhood of $0 \in \r^K$ onto
an open neighborhood of $I \in \mathfrak Q(G)$. Since $\mathfrak Q(G)$ is
connected, the subgroup generated by such a neighborhood is all of
$\mathfrak Q(G)$. It follows that as $\ol{T_\omega}$ commutes with each
$\exp(it_k\ol {B_k})$, it commutes with every element of $\mathfrak
Q(G)$. The unbounded version of Schur's lemma
\cite[(15.12)]{r} then implies that
$\ol{T_\omega} = \lambda I$ for some constant $\lambda$ on
$D(\ol{T_\omega}) = \mathcal H.$ Since
$\ol{T_\omega}$ is the smallest closed extension of ${T_\omega}$ and
$T_\omega \subset T \subset \bar T$, we see that $\bar T =
\lambda I$, whence $T$ itself is a constant multiple of the identity. 
\endproof

\end{section}


\begin{section}{An Algebraic No-Go Theorem}

We first derive an algebraic obstruction to quantization. The idea is
to compare the algebraic structures of Poisson algebras on the one
hand with associative algebras of operators with the commutator
bracket on the other. 
\begin{thm}
\label{thm:algnogo}
Let $\,\cal P$ be a unital Poisson subalgebra of $\,\ci (M)$ or $\ci
(M,\!\bc)$. If as a \la\ $\cal P$ is not commutative, it cannot be
realized as an associative algebra with the commu\-ta\-tor bracket.
\end{thm}


\def\O{{\cal P}}


\noindent {\it Proof.} \: To the contrary, let us assume
that there  is  a  Lie algebra  isomorphism  $\q:\O\rightarrow 
{\mathcal A}$  onto  an  associative algebra ${\mathcal A}$ with the 
commutator bracket. Let us take $m\in M$  and
$f,g\in\O$ such that $\{ f,g\}(m)\ne 0$. In particular, then,
$X_g(m)\ne 0$. Replacing $g$ by $g-g(m)1$, we can assume that
$g(m)=0$. The  Lie subalgebra $\O_m=\{ h\in\O\,|\, X_h(m)=0 \}$   is 
clearly  of  finite codimension in $\O$. Let us  put 
$L=ad^{-1}(\O_m)=\{  h\in\O\,|\,  \{\O,h\}\subset\O_m\}$. Since
$\q(\O_m)$  is  a  finite-codimensional  Lie  subalgebra  of
${\mathcal A}$,  there  is  a  finite-codimensional two-sided
associative  ideal $J$ contained in 
$ad^{-1}(\q(\O_m))=\q(L)$ 
\cite[Prop. 2.1]{Gr2}. But associative ideals are Lie
ideals with  respect to   the   commutator   bracket!   Hence  
$\q^{-1}(J)$    is    a finite-codimensional (say
$(l-2)$-codimensional) Lie ideal of  $\O$ contained  in  $L$.  In 
particular,  some  linear   combination   of
$g^2,g^3,\dots,g^l$, say  $\hat  g=g^k+\sum_{i=k+1}^l a_ig^i,$ $k\ge
2,$ belongs to $\q^{-1}(J)$.  Then  $ad_f^{\,k-2}\hat  g\in \q^{-1}(J)
\subset L$, where
$ad_f \hat g := \{f,\hat g\}$,  and thus
$ad_f^{\,k-1}\hat g = ad_f(ad_f^{\,k-2}\hat g)\in\O_m$. But, as
$g(m)=0$, an easy calculation gives
$$X_{ad_f^{\,k-1}\hat g}(m)= k! \, \{ f,g\}^{k-1}(m)X_g(m)\ne 0,$$ a
contradiction.
\endproof

\ms 

See \cite{Jo} for complementary results regarding $P({\rm h}(2n))$
vis-\`a-vis the Weyl algebra.

In \S 5 we will use this result to prove the nonexistence of
polynomial quantizations of $(P(\fb),\fb)$ when $\fb$ is nilpotent.

\end{section}


\begin{section}{Nilpotent Basic Algebras}

Let $\fb$ be a nilpotent \ba\ on a 
$2n$-dimensional connected symplectic manifold $M$. Since by (B1)
$\fb$ is finitely generated and as every finitely generated nilpotent
Lie algebra is finite-dimensional, 
\cite[Prop. 2]{go98} shows that $M$ must be a coadjoint orbit in
$\fb^*$. Now we have the ``bundlization'' results of Arnal \emph{et
al.}
\cite{ACMP}, Pedersen \cite{Pe}, Vergne \cite{Ve2},
and Wildberger \cite{Wi}, which assert:

\begin{thm}
Let $\fb$ be a \fd\ nilpotent \la. For each
$2n$-dimensional coadjoint orbit
$O \subset \fb^*$, there exists a symplectomorphism {\rm
(``bun\-dl\-ization'')} $\varphi_{O}: T^*\r^{n} \ra O$. We may
consider
$b \in \fb$ as a {\rm (}linear{\rm )} function on
$\fb^*$, and form $\Phi_{O}(b) = b | O \circ \varphi_O.$
Then  cotangent coordinates $(q_1,\ldots,q_n,p_1,\ldots,p_n)$ on
$T^*\r^n$  may be chosen in such a way that $\Phi_O(b)$ has the form
\begin{equation}
\phi_0p_1 + \phi_1(q_1)p_2 + \cdots +
\phi_{n-1}(q_1,\ldots,q_{n-1})p_n + \phi_{n}(q_1,\ldots,q_{n}),
\label{eq:obs} 
\end{equation}

\noi where the $\phi_\alpha$ are polynomials.
\label{thm:bun}
\end{thm}

Thus we may assume that $M = \T \r^n$ and that $\fb$ consists of
elements of the form (\ref{eq:obs}). See \cite{Gra} for an
analogous characterization of transitive nilpotent \la s of
vector fields. 

The canonical example of a nilpotent \ba\ on $\T \r^n$ is the
Heisenberg algebra h$(2n) = \sp _\r\{1,q_\alpha,p_\alpha \,|\, \alpha =
1,\ldots ,n\}.$ It is not difficult to see from (\ref{eq:obs}) that,
up to isomorphism, h(2) is the only nilpotent \emph{basic} algebra on
$\T \r$. This is not true in higher dimensions, however:
\begin{equation*}
\fb = \sp _\r \{1,q_1,p_2,q_1 p_2 + q_2,p_1\}
\label{eq:sfbasis}
\end{equation*}

\noi is a nilpotent \ba\ on $\T \r^2$ which is not isomorphic to h(4).
Regardless, all nilpotent \ba s on $\T \r^n$ enjoy
the following property. We write ${\bf q} =
(q_1,\ldots,q_n)$, etc.
\begin{prop} If $\,\fb$ is a nilpotent basic algebra on $\T \r^n$,
then as \pa s $P(\fb) = \r [{\bf q},{\bf p}].$
\label{prop:poly}
\end{prop}

\noi \emph{Proof.} That $P(\fb) \subseteq \r [{\bf q},{\bf p}]$ is
evident from Theorem~\ref{thm:bun}. The opposite inclusion follows
from an algorithm, developed in \cite[\S 5.4]{Pe}, which constructs
the $\{ q_\alpha,p_\alpha\,|\, \alpha = 1,\ldots, n \}$ as polynomial
functions of elements of a basis of $\fb$. That $P(\fb)$ and $\r
[{\bf q},{\bf p}]$ coincide as Lie algebras is due to the fact that
the bundlization
$\varphi_{O}$ is a symplectomorphism or, equivalently, that the 
coordinates $q_\alpha,p_\alpha$ are
canonical.
\endproof

\ms

We will establish a quantum analogue of this result in the next
section.

Recall
that the central ascending series for $\fb$ is 
\begin{equation*}
\{0\} = \fb^0 \subset \fb^1 \subset \cdots \subset \fb^\ell = \fb
\label{eq:cas}
\end{equation*}

\noi  for some positive integer $\ell$, where $\fb^{s+1} =
ad^{-1}(\fb^s)$. Then $\{\fb,\fb^s\} \subseteq \fb^{s-1}.$ Also note
that $\fb^1$ is the center of $\fb$ which, according to the
transitivity condition in (B3), consists of constants. 
Choose a Jordan-H\"older basis $\{b_1,\ldots,b_K\}$ of $\fb$.
Then $\{b_i,b_j\} = \sum^K_{k =1}c_{ij}^k b_k$, where the
structure constants $c_{ij}^k = 0$ whenever $k \geq \min\{i,j\}$. We
take $b_1 = 1.$ We call the smallest integer $N$ such that $b \in
\fb^{N+1}$ the ``nildegree'' of $b \in \fb$. Then nildeg$(b_i) \leq$
nildeg$(b_j)$ whenever $i < j.$ The nildegree of a
polynomial
$f \in P(\fb)$ is then the smallest integer $N$ such that 
$$\big( ad(b_{i_1})\circ \cdots \circ
ad(b_{i_{N+1}})\big)f = 0$$
for all $i_1,\ldots,i_{N+1} \in \{1,\ldots,K\}.$ 

\end{section}


\begin{section}{Proof of Theorem~\ref{thm:nil} and Related Results}

Before proving Theorem~\ref{thm:nil} we establish several
results which are useful in their own right.

Let the \ba\ $\fb$ be nilpotent. Fix a \lsa\
$\oo$ of $P(\fb)$ containing $\fb.$ Suppose that $\q:\oo \ra {\rm
Op}(D)$ is a quantization of $(\oo,\fb)$ on some invariant dense
domain $D$ in a Hilbert space. 
\begin{prop}
$\q$ is injective.
\label{prop:inj}
\end{prop}

\noi \emph{Proof.} Let $L = \ker \q$; then given $g \in L$, there is a
$k$ such that
$g \in \oo^k$, where $\oo^k$ is the subspace of $\oo$ consisting of 
polynomials of nildegree at most $k$ in the elements of $\fb$.
Consider the adjoint representation of
$\fb$ on $\oo^k \cap L.$ (This makes sense as $L$ is a Lie
ideal.) This is a nilrepresentation, so by Engel's theorem
\cite[\S X.2]{ns} there exists a nonzero element $f \in \oo^k
\cap L$ such that
$\{f,b\} = 0$ for all $b \in
\fb.$ But then transitivity implies that $f$ is a constant, which
contradicts (Q2). Thus $L = \{0\}$.
\endproof

\ms

Thus condition (Q6) is actually redundant in the case of nilpotent
\ba s. 

Let $\A$ be the associative
algebra generated over $\bc$ by $\{\q(b)\,|\,b \in \fb\}$. The
next result generalizes 
Proposition~\ref{prop:poly} to the quantum context.  
\begin{prop}
$\A$ is isomorphic to a Weyl algebra.\footnote{\,Recall that the
Weyl algebra $W(2k)$ is the associative algebra over $\bc$ generated by
$\{z_\alpha,w_\beta\,|\,\alpha,\beta = 1,\ldots,k\}$ and the relations
$[z_\alpha,w_\beta] = -i\delta_{\alpha\beta}$, $[z_\alpha,z_\beta] = 0
= [w_\alpha,w_\beta]$.} 
\label{prop:Weyl}
\end{prop}

\noi \emph{Proof.} 
First we claim that the center of $\A$ is
just ${\bc}I$. Indeed, suppose $[A,\q(b)] 
\linebreak
= 0$ for all
$b \in \fb$. Since by construction every $A \in {\A}$ has an adjoint,
we may decompose $A$ into its symmetric $A_s$  and skew-symmetric
$A_a$ components. Algebraic irreducibility then implies that the
symmetric operators $A_s$ and $iA_a$ are both scalar multiples of the
identity. 

Next let $\psi$ be the homomorphism of the universal
enveloping algebra
\linebreak 
$U(\q(\fb_\bc))$ into $\A$ determined by the
inclusion $\q(\fb_\bc) \hookrightarrow \A$. Then $J = \ker \psi$ 
\linebreak
is a
two-sided ideal in $U(\q(\fb_\bc))$. Clearly, $\psi$ is an epimorphism
and thus
\linebreak
$U(\q(\fb_\bc))/J \approx \A$. 

Since furthermore $\q(\fb_\bc)$
is nilpotent, the desired result now follows from \cite[Thm.
4.7.9]{Di}.
\endproof

\ms


By requiring
$\q$ to be complex linear, we may view it as a quantization of the
complexification $\oo_\bc$. We next
prove that $\q$ maps $\oo_\bc$ into $\A$. That ``polynomials quantize
to polynomials'' can be regarded as a generalized ``\vn\ rule,'' cf.
\cite{go98}.
\begin{prop}
$\q(\oo_\bc) \subseteq \A$.
\label{prop:into}
\end{prop}

\noi \emph{Proof.} We argue inductively on the nildegree of $f \in
\oo$ that $\q(f) \in \A$. In nildegree 0 this follows immediately
from transitivity and (Q2). Now suppose it is also true for
polynomials in $\oo$ of nildegree $J \leq N$, and let $f \in \oo$
have nildegree $N+1$. Then for each $b \in \fb,$
$$\big[\q(f),\q(b)\big] = -i\hbar \q\big(\{f,b\}\big) \in \A$$
by (Q1) and the inductive hypothesis, since 
$\mbox{nildeg}\big(\{f,b\}\big) <
\mbox{nildeg}(f).$ Thus the map
\begin{equation*}
W \mapsto \big[\q(f),W\big]
\label{eq:hdiff}
\end{equation*}

\noi defines a derivation of the associative algebra $\A$. 
As it is well known that every derivation of a Weyl algebra is inner
\cite[\S 4.6.8]{Di}, by Proposition \ref{prop:Weyl} there is thus
an $A \in \A$ such that $\big[\q(f),W\big] = [A,W]$ for all $W
\in \A$. Algebraic irreducibility then implies that the
symmetric operator $\q(f)$ and the symmetric component $A_s$ of $A$ 
differ by a constant multiple of $I$. Thus the inductive step is
proved and so $\q(\oo)$, and hence $\q(\oo_\bc)$, are contained in
$\A$. 
\endproof

\ms 

We are finally ready to show that there is
no quantization of $(P(\fb),\fb)$. Set $B_i=\q(b_i)$. As
$\q(\fb_{\bf C})$ is nilpotent, we may likewise define the nildegree
of the $B_i$ etc.\footnote{\,This is so even though $\q$ need not be a
nilrepresentation.} Since $\q$ is faithful
we have that nildeg$(B_i) =$ nildeg$(b_i)$.

\ms

\noi \emph{Proof of Theorem~\ref{thm:nil}}. Suppose
that $\q: P(\fb) \ra {\rm Op}(D)$ were a quantization of
$(P(\fb),\fb)$. Let
${\mathcal P} = P(\fb)_\bc$. {}From Proposition~\ref{prop:into} we
know that 
$\q({\mathcal P}) \subseteq \A$, and 
from Proposition~\ref{prop:inj} we have that 
$\q$ is injective. Thus if we can show that $\q$ is surjective, then
$\q$ will be a
\la\ isomorphism of $\,\mathcal P$ onto $\A$, thereby
contradicting Theorem~\ref{thm:algnogo}.

To this end, we shall prove inductively that
\begin{description}
\item \rule{0mm}{0mm}
\begin{enumerate}
\vspace{-4ex}
\item[($*_N$)] {\it If the monomial $\,b_1^{\; r_1}\cdots b_K^{\;
r_K}\in P(\fb)$ is of nildegree	$J$, $J \leq N$,	then}
\begin{equation*} 
\hspace{-6ex}
\q(b_1^{\; r_1}\cdots b_K^{\; r_K})  =  \s(B_1^{\; r_1}\cdots
B_K^{\; r_K}) 
 + {\it  \ polynomials \ of\ nildegree } < J,
\end{equation*}
\end{enumerate}\end{description}
\noi where $\s$ denotes symmetrization over all factors.

We have already seen that condition ($*_0$) holds. Now assume that
$b_1^{\; r_1}\cdots b_K^{\; r_K}$ has nildegree $N+1.$ By (Q1),
\vspace{1ex}
\begin{eqnarray*} \big[\q(b_1^{\; r_1}\cdots b_K^{\; r_K}),B_j\big]
& = & -i\hbar \, \q\big(\{b_1^{\; r_1}\cdots b_K^{\;
r_K},b_j\}\big) \nonumber
\\ & = & -i\hbar \, \q\!\left(\sum_{l=1}^K r_l\, b_1^{\; r_1}\cdots
b_l^{\; r_l-1}\{b_l,b_j\}
\cdots b_K^{\; r_K}\right) \nonumber \\
& = & -i\hbar \sum_{l,m =1}^K r_l \,c_{lj}^m \,\q \!\left(b_1^{\; r_1}
\cdots  b_m^{\; r_m+1} \cdots b_l^{\; r_l-1} \cdots b_K^{\;
r_K}\right) \nonumber
\\ & = &  -i\hbar \sum_{l,m =1}^K r_l\, c_{lj}^m\,  \s(B_1^{\; r_1}
\cdots  B_m^{\; r_m+1} \cdots B_l^{\; r_l-1} \cdots B_K^{\; r_K})
\nonumber  \\ & & +  {\it  \ polynomials\ of\ nildegree} < N
\label{eq:induct}
\end{eqnarray*}

\noi where the last equality follows from ($*_N$), since 
\begin{eqnarray*}
\mbox{nildeg}\big(c_{lj}^m\,b_1^{\; r_1} \cdots b_m^{\; r_m+1}
\cdots b_l^{\; r_l-1} \cdots  b_K^{\; r_K}\big)
& \leq  & \mbox{nildeg}\big(\{b_1^{\; r_1}\cdots b_K^{\;
r_K},b_j\}\big) \\
& < &
\mbox{nildeg}\big(b_1^{\; r_1}\cdots b_K^{\; r_K}\big).
\end{eqnarray*} 

\noi Furthermore, direct computation yields
\begin{equation*} \big[\s(B_1^{\; r_1}\cdots B_K^{\;
r_K}),B_j\big] = -i\hbar \sum_{l,m =1}^K r_l\, c_{lj}^m\, \s(B_1^{\;
r_1}\cdots  B_m^{\; r_m+1} \cdots B_l^{\; r_l-1} \cdots B_K^{\; r_K}).
\label{eq:recursion}
\end{equation*}
\vskip -2ex
\noi Consequently for each $j = 1,\ldots,K$,
\begin{equation*}
\big[\q(b_1^{\; r_1}\cdots b_K^{\; r_K})- \s(B_1^{\;
r_1}\cdots B_K^{\; r_K}), B_j\big] = {\it  \ polynomials\ of\
nildegree} < N.
\label{eq:diff}
\end{equation*}
\noi This implies that the polynomial
$\q(b_1^{\; r_1}\cdots b_K^{\; r_K})- \s(B_1^{\; r_1}\cdots B_K^{\;
r_K}) \in \A$ has nildegree at most $N$, and
($*_{N+1}$) follows.

Applying ($*_N$) recursively, we see that as the
$\s(B_1^{\; r_1}\cdots B_K^{\; r_K})$ form a basis for $\A$, $\q$ maps
onto $\A$. 
\endproof

\ms

Even though one cannot quantize \emph{all} of
$P(\fb)$, it is possible to quantize `sufficiently small' \lsa s
thereof (see, e.g. \cite{go-r2n}).
We emphasize that Propositions~\ref{prop:inj}--\ref{prop:into} are
valid in this context. It is an open problem to determine the maximal
quantizable \lsa s of
$P(\fb).$

\end{section}


\begin{section}{Solvable Basic Algebras}

We have shown that there is an obstruction to 
quantizing \sm s with nilpotent \ba s. It
is also known that there is an obstruction to quantizing $\T \!S^1$
with the Euclidean basic algebra e(2), which is solvable \cite{GG1}.
Thus it is natural to wonder if the nilpotent no-go theorem extends to
the solvable case. It turn out that it does
\emph{not}: We now show that there is a polynomial 
quantization of $\T \r_+ = \{(q,p) \in \r^2\,|\,q >0\}$ with the
``affine''
\ba\
\[{\rm a}(1) = {\rm span}_\r\{pq,q^2\}.\] 

Upon writing $x = pq,$ $y = q^2$, the bracket relation becomes
$\{x,y\} = 2y.$ Thus a(1) is the simplest example of a solvable
algebra which is not nilpotent. The corresponding polynomial algebra
$P = \r[x,y]$ is free, and has the crucial feature that for each $k
\geq 0$, the subspaces $P_k$ are \emph{ad}-invariant, i.e.,
\begin{equation}
\{P_1,P_k\} \subset P_k.
\label{ad}
\end{equation}

\noi (Here $P_k$ denotes the subspace of homogeneous polynomials of
degree $k$ in $x$ and $y$, and $P^k = \oplus_{l=0}^k P_l$. Note that
$P_1 = {\rm a}(1)$). Because of this
$\{P_k,P_l\}
\subset P_{k+l-1},$ whence each $P_{(k)} = \oplus_{l \geq k}P_l$ is a
Lie ideal. We thus have the semidirect sum decomposition
\begin{equation}
P = P^1 \ltimes P_{(2)}.
\label{decomp}
\end{equation}

Now on to quantization. In view of (\ref{decomp}), we can obtain
a quantization $\q$ of $P$ simply by finding an
appropriate representation of $P^1 = \r \oplus P_1$ and setting
$\q(P_{(2)}) = \{0\}$!

The connected, simply connected covering group of
a(1) is ${\rm A}(1)_+ = \r \rtimes \r_+$ with the composition law
$$(\nu,\lambda)(\beta,\delta) = (\nu +
\lambda^2\beta,\lambda\delta).$$ (A(1)$_+$ is
isomorphic to the group  of
orientation-preserving affine transformations of the line, whence the
terminology.) Since A(1)$_+$ is a semidirect product we can
generate its unitary representations by induction.
Following the recipe in \cite[\S 17.1]{b-r} we obtain two
one-parameter families of unitary representations $U_{\pm}$ of
A(1)$_+$ on $L^2(\r_+,dq/q)$ given by
\[\big(U_{\pm}(\nu,\lambda)\psi\big)(q) = e^{\pm i\mu \nu
q^2}\psi(\lambda q)\]

\noi with $\mu > 0.$ We identify the parameter $\mu$ with
$\hbar^{-1}$. According to Theorems 4 and 5 in
\cite[\S 17.1]{b-r} the remaining two representations (one for each
choice of sign) are irreducible and inequivalent; moreover, up to
equivalence these are the only nontrivial irreducible ones. 

Let $D \subset L^2(\r_+,dq/q)$ be the linear span of the
functions $\sqrt{q}\,h_k(q)$, where the $h_k$ are the Hermite
functions. Writing
$\pi_{\pm} =-i\hbar\, dU_{\pm}$ we get the 
representation(s) of a(1) on the dense subspace $D$:
\[\pi_{\pm}(pq) = -i\hbar \, q\frac{d}{dq},\;\;\pi_{\pm}(q^2)
= \pm q^2.\]

\noi Extend these to $P^1$ by taking $\pi_{\pm}(1) =
I$, and set $\q_{\pm} = \pi_{\pm} \oplus 0$ (cf. (\ref{decomp})).
Clearly (Q1)--(Q3) hold, by construction (Q4) is satisfied, and
$\q_{\pm}\restriction {\rm a}(1) = \pi_{\pm}$ is faithful. Finally, it
is straightforward to verify that $D$ consists of analytic vectors
for both $\pi_{\pm}(pq)$ and $\pi_{\pm}(q^2)$.  Thus
$\q_{\pm}$ are the required quantization(s) of $(P,P_1)$.

\ms
\noi \emph{Remarks.} 1. The $+$ quantization of a(1) is exactly what
one obtains by geometrically quantizing $\T\r_+$ in the vertical
polarization. Carrying this out, we get $\h = L^2(\r_+,dq)$ and
\[pq \mapsto -i\hbar\left(q\frac{d}{dq} +
\frac{1}{2}\right),\;\;q^2 \mapsto q^2.\]

\noi  The $+$ quantization is unitarily equivalent to this via the
 transformation \linebreak $L^2(\r_+,dq/q) \rightarrow
L^2(\r_+,dq)$ which takes $f(q) \mapsto f(q)/\sqrt{q}.$ 

\vskip 6pt

2. Note that ${\rm a}(1) \subset \mbox{sp}(2,\r)$. In fact, the $+$ 
quantization is equivalent to
the restrictions to a(1) of  the metaplectic
representations of sp$(2,\r)$ on both
$L^2_{\mbox{\scriptsize even}}(\r,dq)$ and
$L^2_{\mbox{\scriptsize odd}}(\r,dq)$ \cite[\S
5.1]{go98}.  

\vskip 6pt

3. Since $\q(P_{(2)}) = 0$, the quantization is somewhat
`trivial.' However, there are quantizations which are nonzero on
$P_{(2)}$: for instance, set $\q(x^k) = k\q(x)$ for $k>0$, $\q(x^ly) =
\q(y)$, and $\q(x^ly^m) = 0$ for $m > 1.$

\vskip 6pt

4. Our quantization of $\T \r_+$ should be contrasted with that
given in \cite[\S 4.5]{Is}. Also, we observe that this example is
symplectomorphic to $\r^2$ with the basic algebra $\sp\{p,e^{2q}\}$.

\vskip 6pt

5. This is not the first example of a polynomial quantization; in
\cite{Go} a quantization of the entire \pa\ of the torus was
constructed. However, the \ba\ in that example was
\emph{infinite}-dimensional.

\ms

What makes this example work? After comparing it with other
examples, it is evident that this polynomial quantization exists
because we cannot decrease
degree \emph{in} $P$ by taking Poisson brackets. (That is, we have
(\ref{ad}) as opposed to merely
$\{P_1,P_k\} \subset P^k.$) Based on this observation, it seems
reasonable to suspect that there is an obstruction to quantizing
$(P(\fb),\fb)$ iff it is possible to lower degree in
$P(\fb)$ by taking \pb s. We shall pursue this line of investigation 
elsewhere (cf. also \cite{go98}).

\ms

We thank M. Gerstenhaber, B. Kaneshige, and N. Wildberger for providing
us with helpful comments and references.

\end{section}



\end{document}